\documentclass[aps,prb,preprint]{revtex4}

\usepackage{graphicx}

\begin{document}


\title{Quantized Anomalous Hall Effect in Magnetic Topological
  Insulators}



\author{Rui Yu,$^{1}$ Wei Zhang,$^{1}$ Hai-Jun Zhang,$^{1,2}$
  Shou-Cheng Zhang,$^{2,3}$ \\ Xi Dai,$^{1\ast}$ Zhong Fang$^{1\ast}$}


\affiliation{$^{1}$Beijing National Laboratory for Condensed Matter
  Physics and Institute of Physics, Chinese Academy of Sciences,
  Beijing 100190, China, \\ $^{2}$Department of Physics, McCullough
  Building, Stanford University Stanford, CA 94305-4045, \\ $^{3}$
  Center for Advanced Study, Tsinghua University, Beijing 100084,
  China\\ }


\date{\today}

\begin{abstract}
  The Hall effect, the anomalous Hall effect and the spin Hall effect
  are fundamental transport processes in solids arising from the
  Lorentz force and the spin-orbit coupling respectively. The quantum
  versions of the Hall effect and the spin Hall effect have been
  discovered in recent years. However, the quantized anomalous Hall
  (QAH) effect has not yet been realized experimentally. In a QAH
  insulator, spontaneous magnetic moments and spin-orbit coupling
  combine to give rise to a topologically non-trivial electronic
  structure, leading to the quantized Hall effect without any external
  magnetic field. In this work, based on state-of-art first principles
  calculations, we predict that the tetradymite semiconductors
  Bi$_2$Te$_3$, Bi$_2$Se$_3$, and Sb$_2$Te$_3$ form magnetically
  ordered insulators when doped with transition metal elements (Cr
  or Fe), in sharp contrast to conventional dilute magnetic
  semiconductor where free carriers are necessary to mediate the
  magnetic coupling. Magnetic order in two-dimensional thin films
  gives rise to a topological electronic structure characterized by a
  finite Chern number, with quantized Hall conductance
  $e^{2}/h$. Experimental realization of the long sought-after QAH
  insulator state could enable robust dissipationless charge transport
  at room temperature.
\end{abstract}

\pacs{}

\maketitle


\section{Introduction}

Soon after the observation of the Hall effect, Hall also observed
the anomalous Hall effect (AHE)~\cite{Hall1880_a,Hall1880_b} in
ferromagnetic metals, where the Hall resistance arises from the
spin-orbit coupling (SOC) between electric current and magnetic
moments, even in the absence of external magnetic field. Recent
progress on the mechanism of AHE have established a link between the
AHE and the topological nature of Hall current due to SOC by
adopting the Berry-phase
concepts~\cite{Nagaosa-AHE,Jungwirth,Fang2003_a,Fang2003_b}, in
close analogy to the intrinsic spin Hall effect~\cite{SHE1,SHE2}.
Given the experimental discovery of the quantum Hall~\cite{Klitzing}
and quantum spin Hall (QSH) effects~\cite{Bernevig2006, Konig2007},
it is natural to ask whether the AHE can also be quantized in a
topological magnetic insulator, without the external magnetic field
and the associated Landau levels.

A simple mechanism for the QAH insulator has been proposed in a
two-band model of a two dimensional (2D) magnetic
insulator~\cite{Qi2006}. In the limit of vanishing spin-orbit
coupling, and large enough exchange splitting, the majority spin
band is completely filled and minority spin band is empty. When the
exchange splitting is reduced, the two bands intersect each other,
leading to a band inversion, say near the $\Gamma$ point. The
degeneracy at the interaction region can be removed by turning on
the SOC, giving rise to an insulator state with a topologically
non-trivial band structure characterized by a finite Chern number
and chiral edge states characteristic of the QAH state (see figures
(1) and (2) of Ref.~\cite{Qi2006}). In alternative mechanisms of
realizing the quantum Hall state without an external magnetic field,
bond currents on a honeycomb lattice has been
proposed~\cite{Haldane}, and the localization of the band electron
could also give rise to a quantized plateau~\cite{QAHE-Nagaosa}.
However, these mechanisms seem harder to realize experimentally.

Therefore, the crucial ingredients for realizing an QAH state are: (1)
a ferromagnetically ordered 2D insulator which breaks the
time-reversal symmetry; (2) a band inversion transition with strong
SOC. In comparison with the QSH state, which satisfies the second
condition~\cite{Bernevig2006, Konig2007}, the first condition of the
ferromagnetic order in an insulating state is harder to realize.
However, the QSH insulators are usually good starting candidates for
finding QAH effect due to its proximity to the band inversion. Given
the QSH effect in HgTe quantum wells~\cite{Bernevig2006, Konig2007}
through a band inversion transition by varying the thickness of the
quantum well, the magnetically doped HgMnTe has been proposed as a
candidate for the QAH insulator~\cite{CXLiu2008}. Unfortunately, the
Mn moments do not order spontaneously in HgMnTe, and an additional,
small Zeeman field is required for the magnetic alignment of the Mn
moments.

Recently, tetradymite semiconductors Bi$_2$Te$_3$, Bi$_2$Se$_3$, and
Sb$_2$Te$_3$ have been theoretically predicted and experimentally
observed to be topological insulators with the bulk band gap as
large as 0.3eV in Bi$_2$Se$_3$~\cite{HJZhang2009, Xia2009,
Chen2009}. The electronic states close to the Fermi level are the
bonding and anti-bonding $p$ orbitals, and these states undergo a
band inversion at the $\Gamma$ point when the strength of the
spin-orbit coupling is increased~\cite{HJZhang2009}. When the
thickness is reduced, the three dimensional (3D) topological
insulator crosses over to a 2D topological insulator in an
oscillatory fashion, as a function of the thickness~\cite{2Dcross}.
In this work, we predict that this family of compounds doped with
proper transition metal elements (Cr or Fe) should give the QAH
state, the long sought-after quantum Hall state without any external
magnetic field. In sharp contrast to the ferromagnetic order in
conventional dilute-magnetic-semiconductors (DMS), where free
carriers are necessary to mediate the ferromagnetic couplings via
the RKKY mechanism~\cite{DMS,Zener,Dietl,Ohno}, the magnetic dopants in tetradymite
compounds naturally order ferromagnetically, mediated by the large
spin susceptibility (of the van Vleck type) of the host insulator
state. Due to the proximity to the band inversion transition, the
exchange splitting leads to a topologically non-trivial insulating
band structure with a finite Chern number. The recent experimental
progresses on this family of compounds have demonstrated that
well-controlled layer-by-layer MBE thin film growth can be
achieved~\cite{Kehui,Xue1,Xue2,Xue3}, and various transition metal
elements (such as Ti, V, Cr, Fe) can be substituted into the parent
compounds with observable ferromagnetism even above
100K~\cite{FM-BiSe_a,FM-BiSe_b,FM-BiSe_c}. By quantitative
first-principles calculations, we further show that such
ferromagnetic QAH insulating state can be also reached with Tc as
high as around 70K.  We will start from simple physical pictures
based on model analysis, and then conclude by quantitative
first-principles calculations.

\section{The Van-Vleck paramagnetism in Bi$_2$Se$_3$ family compounds}

We first discuss minimal requirements for the ferromagnetic insulator
phase in a semiconductor system doped with dilute magnetic ions under the
assumption that the magnetic exchange among local moments is mediated by 
the band electrons.  The
whole system can then be divided into two sub-systems describing the
local moments and band electrons respectively, which are coupled by a
magnetic exchange term.  Therefore if we only consider the spatially
homogeneous phase, the total free energy of the system can thus be
written as,
\[
F_{total}=\frac{1}{2}\chi_{L}^{-1}M_{L}^{2}+\frac{1}{2}
\chi_{e}^{-1}M_{e}^{2}-J_{ex}M_{L}M_{e}-\left(M_{L}+M_{e}\right)H\]
where $\chi_{L/e}$ is the spin susceptibility of the local
moments/electrons, $M_{L/e}$ denotes the magnetization for the local
moment and electron sub-system and $J_{ex}$ is the magnetic exchange
coupling between the local moments and electrons. The onset of the
ferromagnetic phase can be determined when the minimization
procedure of the above free energy gives a non-zero magnetization
without the external magnetic field $H$, which leads to
$J_{ex}^{2}-\chi_{L}^{-1}\chi_{e}^{-1}>0$, or equivalently
$\chi_{L}>\frac{1}{J_{ex}^{2}\chi_{e}}$. In the dilute limit, we can
neglect the direct coupling among the local moments, and then the
spin susceptibility of the local moment sub-system takes the
Curie-Weiss form, $\chi_{L}=x\mu^2_{J}/(3k_BT)$, where $x$ is the
concentration of the magnetic ions and $\mu_J$ is the magnetic
moment of a single magnetic ion.  Therefore, in order to have non-zero
FM transition temperature, the above criterion requires a sizable
spin susceptibility of the electron sub-system $\chi_e$.  In most of
the DMS systems, e.g. (Ga$_{1-x}$Mn$_x$)As, the electronic spin
susceptibility is negligible for the insulator phase and finite
carrier concentration is required to conduct the magnetic coupling
among local moments. As we shall discuss below, unlike the situation
in GaAs, sizable spin susceptibility even exists through the Van
Vleck paramagnetism for the insulator Bi$_2$Se$_3$, providing a new
mechanism for ordering the doped magnetic ions.

The non-zero paramagnetic spin susceptibility for a band insulator has
been known as the Van-Vleck paramagnetism\cite{Van_Vleck} for a long
time. For temperature much less than the band gap, the spin
susceptibility for a band insulator can be obtained by the second
order perturbation on the ground state, which can be written as,
\begin{equation}
\chi^{zz}_{e}=\sum_{E_{nk}<\mu;E_{mk}>\mu}4\mu_0\mu^2_B\frac{\left\langle
  nk\right|\hat{S_{z}}\left|mk\right\rangle \left\langle
  mk\right|\hat{S_{z}}\left|nk\right\rangle }{E_{mk}-E_{nk}}.
\end{equation}
where $\mu_0$ is the vacuum permeability and $\mu_B$ is the Bohr magneton.
It is then quite obvious that the van-Vleck paramagnetism is caused
by the mixing of the conduction and valence bands induced by the
spin operator in the presence of SOC. Such a mechanism is absent in
the GaAs system, because the gap in GaAs is between mostly $s$ and
$p$ bands and the matrix element of $\hat{S_{z}}$ is nearly zero
between conduction and valence bands. In the case of Bi$_2$Se$_3$
family, the situation is completely different: both the conduction
and valence bands are formed by the bonding and anti-bonding $p$
orbitals and the energy gap is opened by the spin-orbital coupling
with band inversion, therefore the matrix element of the spin
operator can be very large.

The Van-Vleck paramagnetism in Bi$_2$Se$_3$ family can be well
understood using the following effective $k\cdot p$ Hamiltonian
derived in ref.~\cite{HJZhang2009},
\begin{equation}
H_{3D}\left(\mathbf{k}\right)=\left[\begin{array}{cccc}
    \mathit{\mathcal{M}}(\mathbf{k}) & A_{xy}k_{-} & 0 &
    A_{z}k_{z}\\ A_{xy}k_{+} & \mathit{-\mathcal{M}}(\mathbf{k}) &
    A_{z}k_{z} & 0\\ 0 & A_{z}k_{z} & \mathit{\mathcal{M}}(\mathbf{k})
    & -A_{xy}k_{+}\\ A_{z}k_{z} & 0 & -A_{xy}k_{-} &
    -\mathit{\mathcal{M}}(\mathbf{k})\end{array}\right]
+\epsilon_{0}\left(\mathbf{k}\right)
\label{BiSe-4band}
\end{equation}
with $k_{\pm}=k_{x}\pm ik_{y}$,
$\epsilon_{0}\left(\mathbf{k}\right)=C+D_{xy}
\left(k_{x}^{2}+k_{y}^{2}\right)+D_{z}k_{z}^{2}$,
$\mathit{\mathcal{M}}(\mathbf{k})=M_{0}+B_{xy}
\left(k_{x}^{2}+k_{y}^{2}\right)+B_{z}k_{z}^{2}$~\cite{Param}, in
the basis of $\left|P_{z}^{+},\uparrow\right\rangle$,
$\left|P_{z}^{-},\downarrow\right\rangle$,
$\left|P_{z}^{+},\downarrow\right\rangle$,
$\left|P_{z}^{-},\uparrow\right\rangle$. The $\pm$ in the basis
denote the even and odd parity states and $\downarrow\uparrow$
indicate the spin. As discussed before in Ref. ~\cite{HJZhang2009},
the four low energy states originate from the bonding and
anti-bonding combinations of Bi and Se $P_z$ orbitals, and the
parity is a good quantum number at the $\Gamma$ point due to the
spatial inversion symmetry. The topological properties of the
Bi$_2$Se$_3$ family can be well explained by the band inversion
between two sets of Kramer's degenerate states with opposite parity
at the $\Gamma$ point. Here we will show that the band inversion not
only gives the non-trivial topological index but also increases the
spin susceptibility dramatically. For simplicity, we focus on a 2D
sheet in the k-space with $k_{z}=0$ (the situation of general $k_z$
is qualitatively similar), where the above effective Hamiltonian
reduces to two diagonal blocks as\cite{Bernevig2006,2Dcross}
\begin{equation}
H_{2D}\left(\mathbf{k}\right)=\left[\begin{array}{cc}
h\left(k\right)
    & 0\\ 0 & h^{*}\left(-k\right)\end{array}\right]
\label{H2D}
\end{equation}
with
$h\left(k\right)=\mathit{\mathcal{M}}(\mathbf{k})\sigma_{z}+A_{xy}\left(k_{x}\sigma_{x}+k_{y}\sigma_{y}\right)+\epsilon_{0}\left(\mathbf{k}\right)$.
The two eigen-states of $h\left(k\right)$ give the occupied and
unoccupied bands respectively, which can be denoted as
$\left|ck\right\rangle $and $\left|vk\right\rangle $. It is then
obvious that the amplitude of the matrix element $\left\langle
ck\right|\hat{S_{z}}\left|vk\right\rangle $ reaches the maximum
value when $\mathit{\mathcal{M}}(\mathbf{k})=0$.  The above
condition can only be satisfied when $M_{0}B_{xy}<0$, which is
exactly the condition for the band inversion. In the normal phase
without inverted bands, the maximum value of $\left\langle
ck\right|\hat{S_{z}}\left|vk\right\rangle $ can not be reached and
the spin susceptibility is less pronounced compared to the system
with band inversion.

To further demonstrate this argument, we perform first-principles
calculations (see ref.~\cite{Method} for method) for the electronic
structures and the spin susceptibility of Bi$_2$Se$_3$ bulk. The
results as function of SOC strength are plotted in Fig. 1(a). When the
relative SOC strength $\lambda/\lambda_0$ is larger than 0.5, the
bands at the $\Gamma$ point are inverted, and the spin susceptibility
starts to increase appreciably around this point.

\section{Mean Field Theory for the Magnetic Topological Insulators}

In the following, we apply a tight binding version of the Zener
model\cite{DMS,Zener,Dietl} to calculate the possible Curie temperature at the mean field
level. Let's first assume that some magnetic local moments are
introduced iso-electronically into the Bi$_2$Se$_3$ compounds
without changing the insulating behavior. We show that ferromagnetic
ordering with $T_c$ as high as 70K can be achieved via the van-Vleck
paramagnetism, without requiring the existence of free carriers as
in conventional DMS systems. The total Hamiltonian of our system can
be written as
\begin{equation}
H_{total}=H_{LDA}+\sum_{I\mu\nu,\alpha=xyz}J_{eff}
\mathbf{S}_{I}^{\alpha}\cdot
\mathbf{s}_{\mu\nu}^{\alpha}c_{I,\mu}^{\dagger}c_{I,\nu}+h.c.\label{Ht}
\end{equation}
with
\[H_{LDA}=\sum_{ij,\mu\nu}t_{ij}^{\mu\nu}c_{i,\mu}^{\dagger}c_{j,\nu}\]
is the single particle Hamiltonian (obtained from first-principles
calculations) expressed in the basis of local Wannier
functions~\cite{Method}. Here $i$ and $j$ denote the sites, $I$ only
runs over all the unit cells containing magnetic ions and $\mu\nu$
are the orbital/spin indices. The coupling strength between the
local spin $\mathbf{S}_I$ and the band spin $\mathbf{s}_{\mu\nu}$ of
band electrons is described by the effective exchange parameter
$J_{eff}$.

Performing the mean field approximation and applying the virtual
crystal approximation (CPA) to average out the spatial inhomogeneity,
Eq.(\ref{Ht}) can be written as
\begin{equation}
H_{MF}=H_{e}+H_{L}-xJ_{eff}\mathbf{M}_{L}^{\alpha}
\cdot\mathbf{M}_{e}^{\alpha}
\end{equation}
where $\mathbf{M}_{L}^{\alpha}=\left\langle
\mathbf{S}_{I=0}^{\alpha}\right\rangle $,
$\mathbf{M}_{e}^{\alpha}=\left\langle
\mathbf{s}_{\mu\nu}^{\alpha}c_{0,\mu}^{\dagger}c_{0,\nu}\right\rangle
$ are the mean field magnetization for the local moments and band
electrons respectively, with
\[H_{e}=H_{LDA}+\sum_{i\mu\nu}^{\alpha=xyz}J_{eff}\mathbf{M}_{L}^{\alpha}\cdot
s_{\mu\nu}^{\alpha}c_{i,\mu}^{\dagger}c_{i,\nu}+h.c.\]
and
\[H_{L}=\sum_{I}^{\alpha=xyz}J_{eff}\mathbf{S}_{I}^{\alpha}\cdot
\mathbf{M}_{e}^{\alpha}\]

\begin{figure}[tbp]
\includegraphics[clip,width=7cm,angle=270]{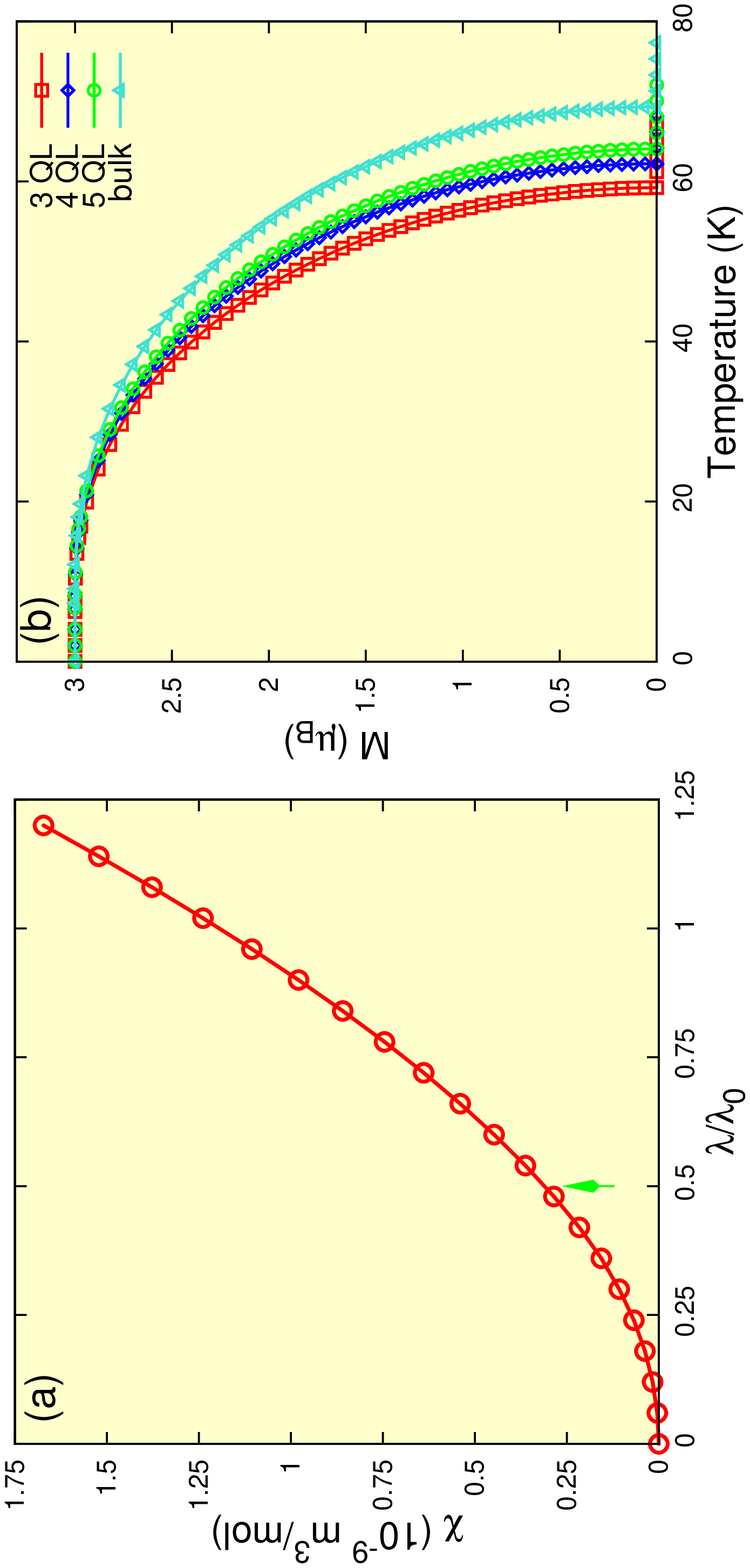}
\caption{\label{fig: bise_spin_sus_and_TC} \textbf{The calculated spin
    susceptibility and Curie temperature for Bi$_{2}$Se$_{3}$}. (a)
  The Van-Vleck type spin susceptibility of Bi$_{2}$Se$_{3}$ bulk as a
  function of spin-orbital-coupling (SOC) strength. $\lambda_{0}$ is
  the real SOC strength, and it is uniformly scaled to simulate
  various SOC strength $\lambda$. When $\lambda/\lambda_{0}$ is larger
  than 0.5, the energy bands at $\Gamma$ are inverted, and
  Bi$_{2}$Se$_{3}$ becomes strong topological insulator. (b) The
  spontaneous magnetization $M$ as a function of temperature for
  Cr-doped ($S_z$=3/2) Bi$_{2}$Se$_{3}$ bulk as well as thin films
  with thickness of 3, 4 and 5 quintuple layers (QL).}
\end{figure}

We choose the concentration of the magnetic ion to be $5\%$, the total
angular momentum of a single magnetic ion to be $J=3/2$ and the
effective exchange coupling $J_{eff}=2.0eV$, which are in the same
order with the quantitative first-principles calculations (as
described later).  The above mean field equations have been solved
self-consistently, and the ordered moment as well as Curie temperature
can be calculated.  In Fig.\ref{fig: bise_spin_sus_and_TC}(b), the
calculated magnetization versus temperature are plotted for both bulk
material and thin film system with 3, 4, 5 quintuple layers (QL)
thicknesses. The Curie temperature for the bulk system is around 70K,
which can be well compared with previous experimental
studies~\cite{FM-BiSe_a,FM-BiSe_b,FM-BiSe_c}. The ferromagnetic
ordering is anisotropic, and the Curie transition is of the Ising
type, with a finite transition temperature in 2D. Our mean field
calculation also indicates that for the thin film system, the Curie
temperature is only slightly reduced due to the finite size effect,
which give rise to the possible QAHE as discussed below.

\section{Electronic structures of Bi$_2$Se$_3$ doped with
transition metal elements}

In this section, we shall show, by parameter-free first principles
calculations, that the insulating magnetic ground state discussed
above can be indeed obtained by proper choice of magnetic dopants.  It
has been suggested by experiments~\cite{FM-BiSe_a,FM-BiSe_b,FM-BiSe_c}
that the magnetic dopants, such as Ti, V, Cr and Fe, will mostly
substitute the Bi ions (or the Sb ions in the case of Sb$_2$Te$_3$),
we therefore concentrate on this situation in the following
discussions. Since the nominal valence of Bi (or Sb) ions are 3+, a
general rule is to find a transition metal element which may have a
stable 3+ chemical state, so that no free carriers are introduced by
this iso-electronic substitution. In addition, due to the coexistence
of orbital and spin degrees of freedom of magnetic dopants (due to the
partially filled $d$-shells), we need a mechanism to quench these
degrees of freedom and stabilize the insulating state. We shall show
that such conditions can be satisfied by the combination of crystal
field splitting and large Hund's rule coupling for Cr and Fe dopants.

We performed self-consistent first-principle
calculations~\cite{Method} for Bi$_2$Se$_3$ doped with various
transition metal elements, Ti, V, Cr and Fe. We use a large supercell
(of size 3$\times$3$\times$1) with one of the Bi atoms substituted by
a magnetic dopant. The crystal structure and the magnetic state are
fully optimized to obtain the ground state. The results shown in Fig.2
suggest that a insulating magnetic state is obtained for Cr or Fe
doping, while the states are metallic for Ti or V doping cases.  To
understand the results, we first point out that, for all the cases,
the dopants are nearly in the 3+ valence state, and we always obtain
the high-spin state due to the large Hund's rule coupling of $3d$
transition metal ions.  Thus the charge and spin degrees of freedom
are nearly frozen. This will directly lead to the insulating state of
Fe-doped samples, because the Fe$^{3+}$ has five $3d$ electrons,
favoring the $d^{5\uparrow}d^{0\downarrow}$ configuration in a
high-spin state, resulting in a gap between the majority and minority
spins. We now consider the orbitals for other dopants. The local
environment of dopants, which substitute the Bi sites, is a octahedral
formed by six nearest neighboring Se$^{2-}$ ions. Such a local crystal
field splits the $d$-shell into $t_{2g}$ and $e_g$ manifolds. This
splitting is large enough to stabilize the
$t_{2g}^{3\uparrow}e_{g}^{0\uparrow}t_{2g}^{0\downarrow}e_{g}^{0\downarrow}$
configuration of Cr$^{3+}$ ion, resulting in a gap between the
$t_{2g}$ and $e_g$ manifolds. For the case of Ti or V doping, even the
$t_{2g}$ manifold is partially occupied, leading to the metallic
state.  It is important to note that, although the local density
approximation in the density functional theory may underestimate the
electron correlation effects, the inclusion of electron-electron
interaction $U$ (such as in LDA+$U$ method) should further enhanced
the gap.

The energy gain due to the spin polarization is about 0.9eV/Fe,
1.5eV/Cr, 0.7eV/V and 0.02eV/Ti, respectively, which are very large
numbers except for the case of Ti substitution. From the spin
splitting of $p$ orbitals of the band electrons, we can estimate the
effective exchange coupling $J_{eff}$ between the local moments and
the $p$ electrons~\cite{J_cal}. The estimated value is around 2.7eV
for Cr and 2.8eV for Fe in Bi$_2$Se$_3$. This exchange coupling is
comparable to that in GaMnAs~\cite{DMS,J_cal}.

\section{The quantized anomalous Hall effect}

Once the ferromagnetically ordered insulating state is achieved with
a reasonable $T_c$, here in this section, we show that QAH effect
can be realized for the 2D thin films.  In the thick slab geometry,
the spatially-separated two pairs of surface states are well defined
for the top and bottom surfaces respectively.  However, with the
reduction of the film thickness, quantum tunneling between the top
and bottom surfaces becomes more and more pronounced, giving rise to
finite mass term in the effective 2D model\cite{2Dcross}. In the
basis of $\left|+\uparrow\right\rangle $
,$\left|-\downarrow\right\rangle $ ,$\left|+\downarrow\right\rangle
$ ,$\left|-\uparrow\right\rangle $ , with $\pm$ referring to the
bonding and anti-bonding combinations of the top and bottom surface
states, and $\downarrow\uparrow$ indicating the spin, the low energy
effective model for the thin film can be written in the form of
$H_{2D}$ as in Eq. \ref{H2D}, with
$h(k)=\Delta_{k}\sigma_{z}+v_{F}\left(k_{x}\sigma_{y}-k_{y}\sigma_{x}\right)$,
$\Delta_{k}=\Delta_{0}+Bk_{\parallel}^{2}$, similar to the
Bernevig-Hughes-Zhang model describing the low energy physics in
HgTe/CdTe quantum well\cite{Bernevig2006}. When $\Delta_{0}B<0$,
band inversion occurs, and the system will be in the QSH phase if
this is the only band inversion between two subbands with opposite
parity, which is exactly the situation in HgTe/CdTe quantum well
system. The situation for Bi$_2$Se$_3$ film is somewhat different in
the sense that more pairs of subbands can get inverted at the
$\Gamma$ point, and the system may still remain in the topologically
trivial state if the band inversion occurs even number of
times~\cite{2Dcross}. Nevertheless, as we shall show below,
regardless of whether the 4-bands system is originally in the
topologically non-trivial phase or not, a strong enough exchange
field will induce QAH effect in this system.

The exchange coupling term induced by the finite magnetization in
the FM phase can be written as,
\begin{equation}
H_{exchange}=\left[\begin{array}{cc} g_{eff}M\sigma_{z} & 0\\ 0 &
    -g_{eff}M\sigma_{z}\end{array}\right]\label{exchange}
\end{equation}
with $M$ denoting the strength of the exchange field and $g_{eff}$
being the effective g-factor of the surface states. Added to the 2D
model (\ref{H2D}), the exchange field increases the mass term of the
upper block and reduces it for the lower block, breaking the time
reversal symmetry. More importantly, a sufficiently large exchange
field can change the Chern number of one of the two blocks.  As
schematically illustrated in Fig.3, if the 4-band system is originally
in the topologically trivial phase, the exchange field will induce a
band inversion in the upper block and push the two sub-bands in the
lower block even farther away from each other.  Therefore the 2D model
with a negative mass in the upper block contributes $e^{2}/h$ for the
Hall conductance. On the other hand, if the system is originally in
the topologically non-trivial phase, both blocks have inverted band
structures. In this case, a sufficiently large exchange field can
increase the band inversion in the upper block and revert the band
inversion in the lower block.  Again the negative mass in the upper
block contributes $e^{2}/h$ for the Hall conductance. These two
scenarios are illustrated in Fig.\ref{fig:chemata}.

To be realistic, we carried out quantitative calculations for
Bi$_2$Se$_3$ films, based on the density functional theory and
linear response formalism~\cite{Method}. A spacial uniform exchange
field is included to take into account the effect of magnetization
in the FM state at the mean field level. In Fig.\ref{fig:
bise_m_gap}(a), (b) and (c), we plot the lowest four sub-band levels
at $\Gamma$ point as a function of exchange field.  The level
crossings (indicated by arrows) between the lowest conduction bands
(blue lines) and valence bands (red lines) are found in all the
three Bi$_2$Se$_3$ thin film systems with different layer thickness.
The corresponding Hall conductance can be obtained using the
following Kubo formula\cite{Nagaosa-AHE},
\begin{equation}
\sigma_{xy}={e^{2}}{\hbar}\sum_{nmk}\frac{Im\left[\left\langle
    nk\right|\hat{v}_{x}\left|mk\right\rangle \left\langle
    mk\right|\hat{v}_{y}\left|nk\right\rangle
    \right]}{\left(E_{nk}-E_{mk}\right)^{2}}
    \left(n_{f}(E_{nk})-n_{f}(E_{mk})\right)
\label{kubo_sxy}
\end{equation}
with $E_{nk}$ being the dispersion of the quantum well sub-bands and
$\hat{v}_{x/y}$ being the velocity operators. In the insulator case,
where the chemical potential is located inside the energy gap between
conduction and valence sub-bands, the Hall conductance is determined
by the first Chern number of the occupied bands and must be an exact
integer in the unit of $e^2/h$. The calculated Hall conductance for
Bi$_2$Se$_3$ thin film with three typical thickness are plotted in
figure \ref{fig: bise_m_gap} (d) as function of exchange field, where
a jump from "0" to "1" in the Hall conductance are observed at the
corresponding critical exchange field for the level crossing.

In Fig.\ref{fig:bise_hall_Platform}, we plot the sub-band dispersion
and corresponding Hall conductance as function of chemical potential
for 3 and 5 QL Bi$_{2}$Se$_3$ thin films with different exchange
fields. We can find that the change induced by the exchange field is
mainly in the area near $\Gamma$ point. For the QAH phases ((d) and
(h)), non-zero integer plateau is observed when the chemical
potential falls into the energy gap. The difference between the 3
and 5 QL cases suggests that the QAH plateau is much narrower for
thick film due to the narrower energy gap coming from the
intersection of top and bottom surface states.

In conclusion we have shown that magnetic dopants Cr and Fe form long
range magnetic order in the insulating state of the Bi$_2$Se$_3$
family of topological insulators. The ferromagnetic order lead to a
topologically non-trivial electronic structure with quantized Hall
conductance without any external magnetic field. In real samples, a
small amount of bulk carriers could always be present, however, if the
concentration is low enough, they will be localized by disorder in two
dimensions, and will not affect the precise quantization of the Hall
plateau. The edge states of our QAH state conducts charge current
without any dissipation in the absence of any external magnetic
field. This effect could enable a new generation of low power
electronic devices.

\begin{acknowledgments}
 We acknowledge the valuable discussions with C. X. Liu,
  X. L. Qi, Q. Niu, S. Q. Shen, and the supports from the NSF of
  China, the 973 Program of China (No. 2007CB925000 and 2010CB923000),
  and the International Science and Technology Cooperation Program of
  China.  SCZ is supported by the US NSF under grant numbers
  DMR-0904264.
\end{acknowledgments}


\clearpage

\begin{figure}[tbp]
\includegraphics[clip,angle=270,width=14cm]{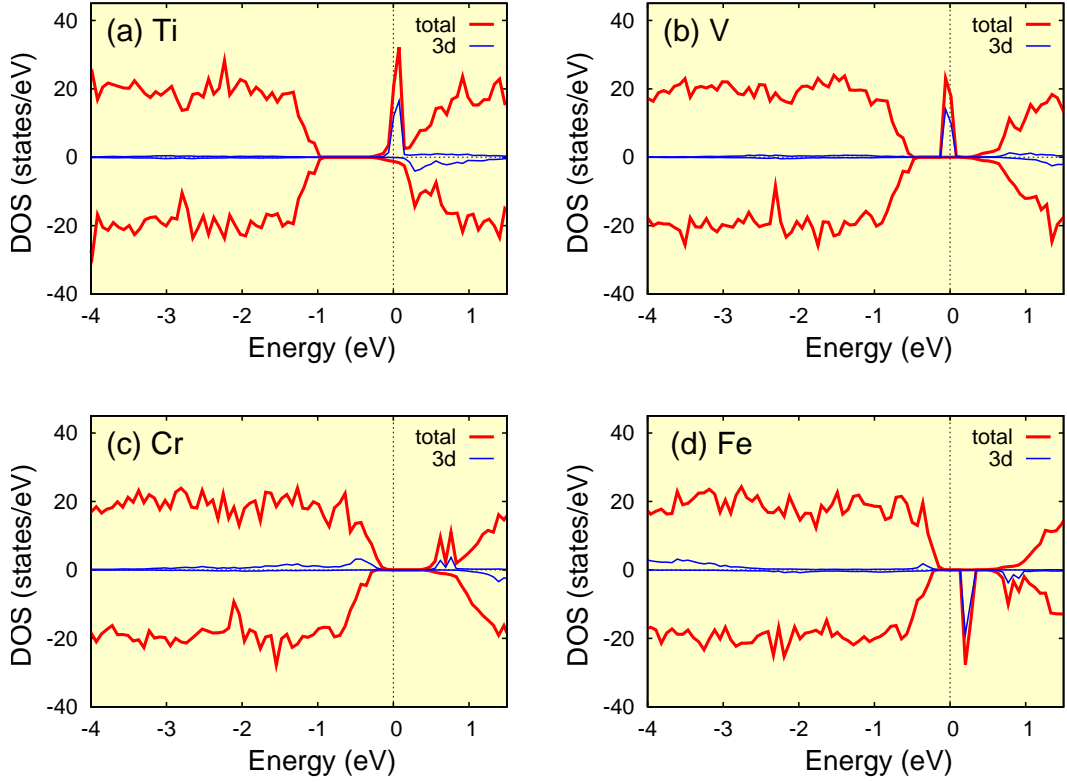}
\caption{\textbf{The calculated density of states (DOS) for
    Bi$_{2}$Se$_{3}$ doped with different transition metal
    elements}. The Fermi level is located at energy zero, and the
  positive and negative values of DOS are used for up and down spin,
  respectively. The blue lines are projected partial DOS of the $3d$
  states of transition metal ions. It is shown that Cr or Fe doping
  will give the insulating magnetic state.}
\end{figure}

\begin{figure}[tbp]
\includegraphics[clip,angle=270,scale=0.6]{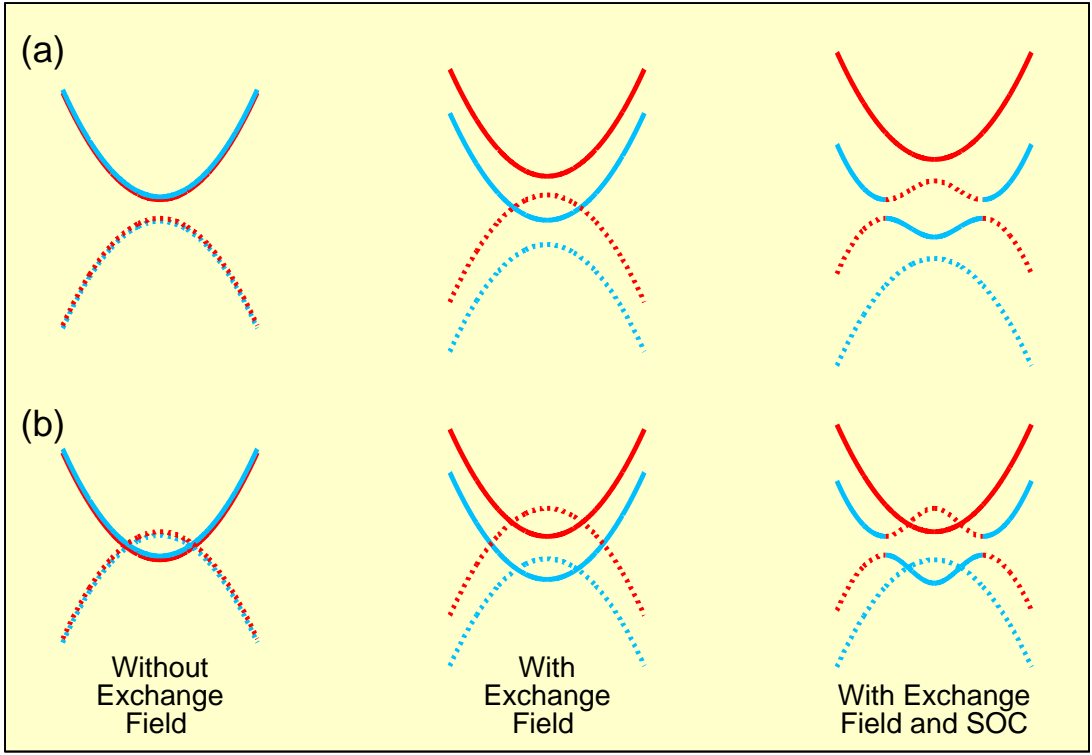}
\caption{\label{fig:chemata}\textbf{Evolution of the sub-bands
    structure upon increasing the exchange field.} The solid (dashed)
  lines denote the sub-band which have even (odd) parity at $\Gamma$
  point, and the blue (red) color denotes the spin down (up)
  electrons.  (a) The initial sub-bands are not inverted. When a
  exchange field is introduced, the spin down sub-bands (blue lines)
  are pushed down and the spin up sub-bands (red lines) are lifted up.
  When the exchange field is strong enough, a pair of inverted
  sub-bands appear (the red dash line and the blue solid line). (b)
  The initial sub-bands are already inverted at $\Gamma$ point. The
  exchange field reverts the band inversion in one pair of sub-bands
  (the red solid line and the blue dashed line) and increase the band
  inversion in the other pair of sub-bands (the red dash line and the
  blue solid line). The SOC opens a gap at the cross points of spin-up
  and spin-down subbands as shown on the right most panel of (a) and
  (b), leading to the QAH state by the mechanism proposed in
  Ref.~\cite{Qi2006,CXLiu2008}}
\end{figure}

\begin{figure}[tbp]
\begin{centering}
\includegraphics[scale=0.6,angle=270]{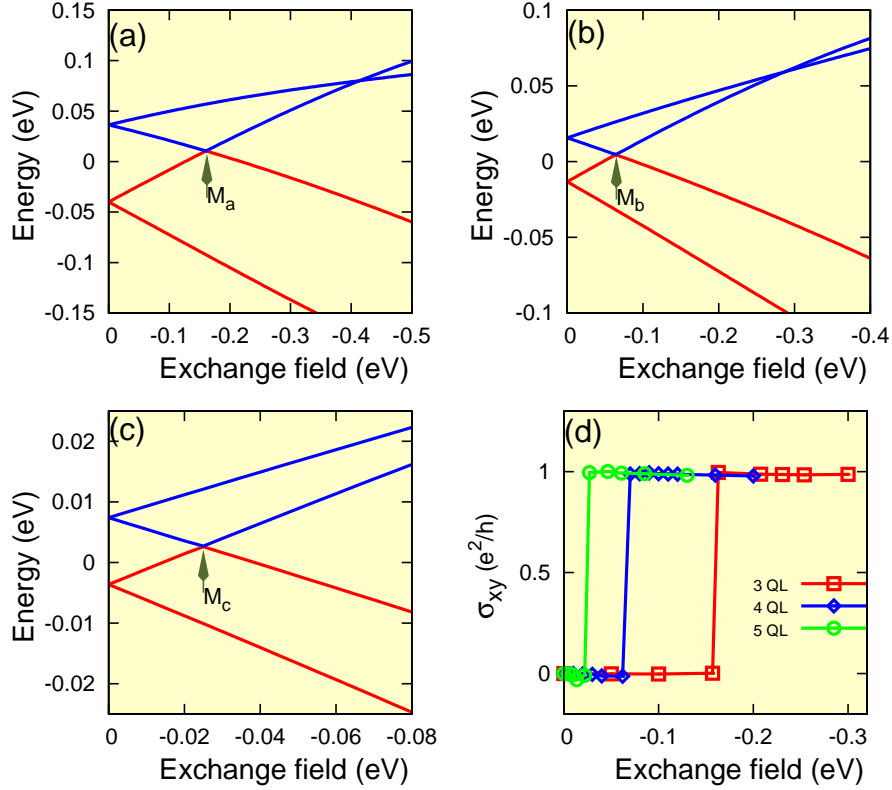}
\par\end{centering}
\caption{\label{fig: bise_m_gap}\textbf{The quantized anomalous
    Hall(QAH) conductance.} (a),\textbf{ }(b) and (c) are the lowest
  sub-bands at the $\Gamma$ point plotted versus the exchange field,
  for Bi$_{2}$Se$_{3}$ films with thickness of 3, 4 and 5QL,
  respectively. The red (blue) lines denote the occupied (unoccupied)
  states. With the increasing of the exchange field, a switch of
  occupied and unoccupied states happens (i.e, a quantum phase
  transition to the QAH phase) at the critical field strength
  indicated by arrows. (d), The calculated Hall conductance using the
  Hamiltonian from first-principles calculations under an uniform
  exchange field (see text for the detail) for 3, 4 and 5QL
  Bi$_{2}$Se$_{3}$ films.  As we expect, the Hall conductance is zero
  before the QAH transition, but $\sigma_{xy}=e^{2}/h$ afterwards.}
\end{figure}

\begin{figure}[tbp]
\begin{centering}
\includegraphics[clip,width=16cm,angle=270]{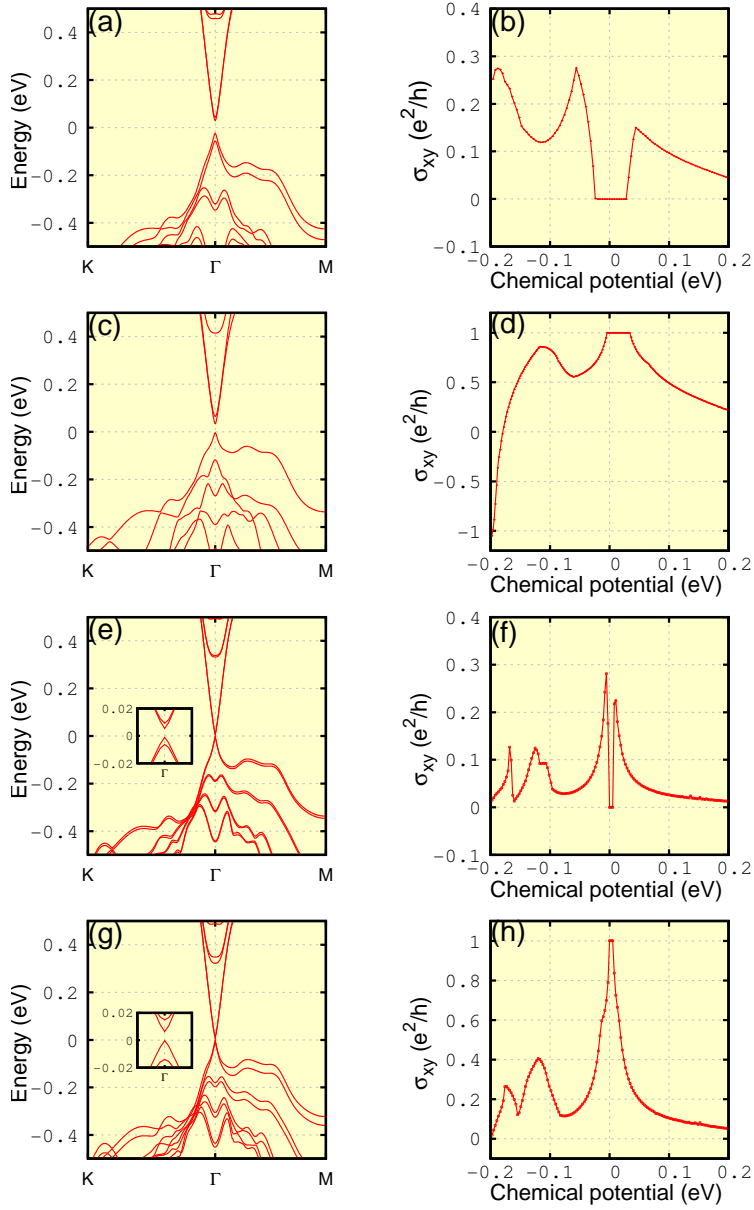}
\par\end{centering}
\caption{\label{fig:bise_hall_Platform} \textbf{The sub-bands
    dispersion and the Hall conductance.} The sub-bands dispersion of
  2D Bi$_{2}$Se$_{3}$ thin film are plotted for (a) 3QL slab at
  exchange field=-0.05eV, (c) 3QL slab at exchange field=-0.24eV, (e)
  5QL slab at exchange field=-0.01eV, and (g) 5QL slab at exchange
  field=-0.04eV. The insets of (e) and (g) zoom out the dispersion
  near $\Gamma$ point. The corresponding Hall conductance are plotted
  in the right panels ((b), (d), (f) and (g)) as the function of the
  chemical potential. The systems in (a) and (e) are normal
  insulators, which have zero platforms in Hall conductance when the
  chemical potential locates in the energy gap. The system in (c) and
  (g) are QAH insulators, in which the Hall conductance have quantized
  value: $e^{2}/h$.}
\end{figure}

\end{document}